\documentclass[preprint, showpacs, superscriptaddress, aps, pre]{revtex4-1}
\usepackage{amsmath}
\usepackage{graphicx}
\usepackage{xcolor}
\usepackage{amssymb}
\usepackage{hyperref}
\usepackage{verbatim}
\usepackage{bm}
\usepackage{float}

\newcommand{\be}{\begin{equation}}
\newcommand{\ee}{\end{equation}}

\begin{document}

\title{Prediction in a driven-dissipative system displaying a continuous phase transition}

\author{Chon-Kit Pun}
\affiliation{Department of Physics, Boston University, Boston, Massachusetts 02215, USA}
\author{Sakib Matin}
\affiliation{Department of Physics, Boston University, Boston, Massachusetts 02215, USA}
\author{W.\ Klein}
\affiliation{Department of Physics, Boston University, Boston, Massachusetts 02215, USA}
\affiliation{Center for Computational Science, Boston University, Boston, Massachusetts 02215, USA}
\author{Harvey Gould}
\affiliation{Department of Physics, Boston University, Boston, Massachusetts 02215, USA}
\affiliation{Department of Physics, Clark University, Worcester, Massachusetts 01610, USA}
\date{\today}

\begin{abstract}
Prediction in complex systems at criticality is believed to be very difficult, if not impossible. Of particular interest is whether earthquakes, whose distribution follows a power law (Gutenberg-Richter) distribution, are in principle unpredictable. We study the predictability of event sizes in the Olmai-Feder-Christensen model at different proximities to criticality using a convolutional neural network. The distribution of event sizes satisfies a power law with a cutoff for large events. We find that prediction decreases as criticality is approached and that prediction is possible only for large, non-scaling events. Our results suggest that earthquake faults that satisfy Gutenberg-Richter scaling are difficult to forecast.

\end{abstract}

\maketitle

\section{Introduction}
A subclass of driven-dissipative systems modeled by a two-dimensional cellular automaton has been proposed to understand the existence of power laws in many complex systems. Examples include the Bak-Tang-Wiesenfeld sandpile model~\cite{BTW_1987}, the Rundle-Jackson model~\cite{Rundle_1977} and the Olami-Feder-Christensen (OFC) model~\cite{OFC_1992}; the latter two models have been used to gain insight into the nature of earthquakes. Many earthquake fault systems display a power-law event size distribution spanning many orders of magnitude. Such a power-law distribution is known as Gutenberg-Richter scaling in the seismology literature~\cite{GR_1944}. For example, the Gutenberg-Richter scaling in Southern California (1984--2000) spans about six orders of magnitude~\cite{Bak_2002}.

There has been a substantial interest in forecasting or predicting earthquakes. However, it has been conjectured that systems at criticality are inherently unpredictable~\cite{BTW_1987}. That is, events of different sizes that satisfy a scale-free distribution are due to the same physical mechanism, and thus there are no distinct precursors to distinguish one event from another~\cite{Bak_1989}. The idea of unpredictability at criticality has been challenged over the years. One school of thought~\cite{Laherrere_1999} has proposed that very large events are due to inherently different mechanisms such as self-reinforcement, synchronization~\cite{Sornette_2009} and nucleation~\cite{Klein_2000}, and are thus in principle distinguishable from smaller events. Some support for this proposal is the use of a technique called the log-periodic-power-law fitting procedure; it has been shown to successfully predict large, non-power law events, such as ruptures in materials~\cite{Anifrani_1995,Johansen_2000} and the end of financial bubbles~\cite{Feigenbaum_1996,Sornette_1996}. 

In this paper we address the question of predictability near and at criticality by applying machine learning to the OFC model. Previous work~\cite{Pepke_1994} has shown that predictability in the OFC model decreases as the conservative limit (a critical point) is approached. We find results consistent with Ref.~\onlinecite{Pepke_1994} and investigate the predictability of events near another critical point in the OFC model: the recently observed noise transition critical point~\cite{Matin_draft}. By using a convolutional neural network (CNN), we find that the event sizes are more difficult to forecast as the critical point is approached and  that only large events that do not satisfy power law scaling can be successfully predicted.

\section{Criticality in the OFC model}
The Olami-Feder-Christensen (OFC) model~\cite{OFC_1992} is a modified version of the spring-block model first proposed by Rundle and Jackson~\cite{Rundle_1977}, which is a simplification of the Burridge-Knopoff model~\cite{Burridge_1967}. The nearest-neighbor OFC model that we will consider consists of a two-dimensional lattice of linear dimension $L$ with each site initially assigned a random value of stress $\sigma$ between the mean residual $\sigma_R$ and the failure threshold $\sigma_F$. We denote the stress on each site, the stress grid, by the vector $\vec{\sigma}=(\sigma_1,\ldots,\sigma_{N=L\times L})$. The system is then driven so that one site reaches $\sigma_F$, a procedure known as the zero velocity limit~\cite{Rundle_1977}. This site is said to fail, and its stress is reduced to $\sigma_R \pm \eta r$, where $\eta$ is the magnitude of the noise and $r$ is a uniform random variable in the range $[-1,1]$. A failing site with stress $\sigma$ distributes stress $(1-\alpha)(\sigma-\sigma_R\mp\eta r)/4$ to its four nearest-neighbor sites, where $\alpha$ is the dissipation parameter. The failure of one site can trigger other sites to fail, thus creating an avalanche. The avalanche or event stops when the stress of all sites is less than $\sigma_F$. We denote the number of failing sites, or the size of the event, by $s$. The system is then driven again using the zero velocity limit. 

The OFC model is believed to approach criticality in the conservative limit $\alpha \to 0$~\cite{Christensen_1991,Serino_2001}. Recently, it has been found that even for $\alpha>0$, there exists a phase transition at a critical value of the noise $\eta_c\approx 0.07$~\cite{Matin_draft}. This phase transition is characterized by the event size distribution $n_{s}$ of the form
\be
n_s \sim s^{-\tau} \exp(-(s/s_c)^{\sigma}) \label{n_s}
\ee
with
\be
s_c = (\eta-\eta_c)^{-1/\sigma},
\ee
and $\tau=1.04 \pm 0.14$ and $\sigma=0.43 \pm 0.03$.
The mean cluster size $\chi$ diverges as $(\eta-\eta_c)_+^{-\gamma}$ and the connectedness length diverges as $(\eta-\eta_c)_+^{-\nu}$~\cite{Stauffer_1979} with $\gamma = 2.01 \pm 0.14$ and $\nu= 1.20 \pm 0.13$, consistent with the scaling relations $\gamma=(3-\tau)/\sigma$ and $\nu=(\tau-1)/d\sigma$, where $d=2$ is the spatial dimension. Note that $n_s$ satisfies power law scaling for $s \lesssim s_c$~\cite{Matin_draft}.

\section{Supervised machine learning}
\begin{figure}[t]
\includegraphics[width=7cm]{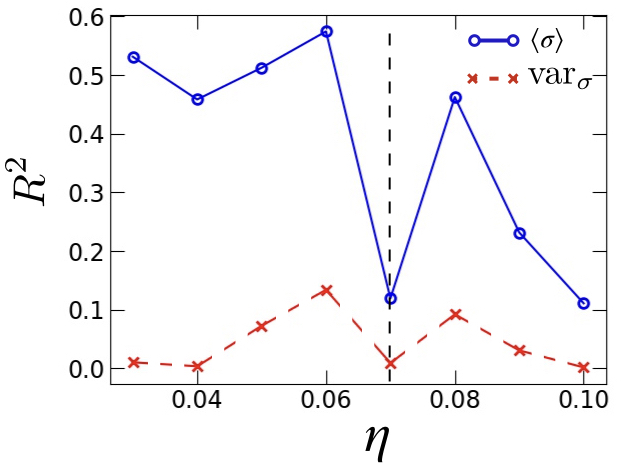}
\caption{\label{fig:R2_noise} The coefficient of determination $R^2$ of the correlations between the true event size $s$ and the average stress $\langle \sigma \rangle$ ($\color{blue} \circ$) and the correlation between $s$ and the variance of the stress ($\color{red} \times$) at different values of the noise $\eta$. The vertical dashed line indicates the critical noise $\eta_c\approx 0.07$~\cite{Matin_draft}. Note that there is a significant correlation between the average stress and the event size for $\eta<\eta_c$. Correlations between the event size and the spatial variance of the stress are insignificant for all values of the noise.}
\end{figure}

Our goal is to predict the event size (the number of failed sites) given the stress grid before stress has been added using the zero velocity limit and before the onset of an event. Figure~\ref{fig:R2_noise} shows that the event size $s$ is strongly correlated with the average stress for $\eta<\eta_c$ and weakly correlated for $\eta \ge \eta_c$. The event size is weakly correlated with the variance of the stress for all values of $\eta$.

To force the CNN to learn higher order features, we first remove the correlations between the event size $s$ and the first and second moments of the stress grid. We normalize each stress grid $\vec{\sigma}$ by its average stress and the spatial variance. That is, we rescale the stress $\sigma_{i,\mu}$ at site $i$ for sample $\mu$ to $\tilde{\sigma}_{i,\mu} \equiv (\sigma_{i,\mu}-\langle{\sigma_{\mu}}\rangle)/\sqrt{\mbox{var}_{\mu}}$, where $\langle \sigma_\mu\rangle=1/N\sum_{i=1}^N \sigma_{i,\mu}$ is the mean stress per site of sample $\mu$ and $\mbox{var}_{\mu} \equiv \sum_{i=1}^N(\sigma_{i,\mu}-\langle \sigma_\mu \rangle)^2/N$ is the spatial variance of the stress. In the following all references to the stress will be to the rescaled stress and we will omit the tilde symbol. We will train the CNN regressor using the rescaled stress grid $\vec{\sigma}$, sampled with quasi--uniform event sizes (see the appendix).

\begin{figure}[b]
\centering
\includegraphics[width=15cm]{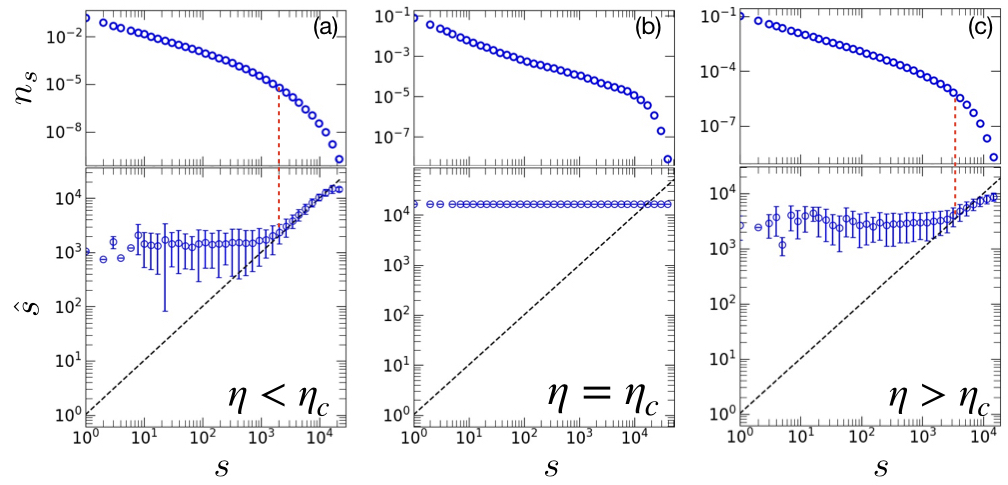}
\caption{The event size distribution $n_s$ versus $s$ (top row) and the true values of $s$ versus the predicted $\hat{s}$ event sizes for (a) $\eta<\eta_c$, (b) $\eta=\eta_c$, and (c) $\eta>\eta_c$ (bottom row). Perfect prediction is represented by the dashed diagonal line. Note that the CNN successfully predicts event sizes only for $s \gtrsim s_c$. The vertical dotted line denotes $s=s_c$.}
\label{fig:YpredictYtest}
\end{figure}

To assess the performance of the machine, we show in Fig.~\ref{fig:YpredictYtest} the predicted event size $\hat{s}$ versus the true event size $s$. The top row shows the event size distribution $n_s$ at different values of $\eta$. The bottom row shows the predicted event sizes  versus the true event sizes. We see that for $\eta < \eta_c$, the machine performs impressively at predicting events that are larger than $s_c$ [see Fig.~\ref{fig:YpredictYtest}(a)]. For $\eta > \eta_c$, the machine performs less impressively [see Fig.~\ref{fig:YpredictYtest}(c)]. At $\eta = \eta_c$, the machine fails at predicting events of all sizes [Fig.~\ref{fig:YpredictYtest}(b)]. 

\begin{figure}[t]
\centering
\includegraphics[width=7cm]{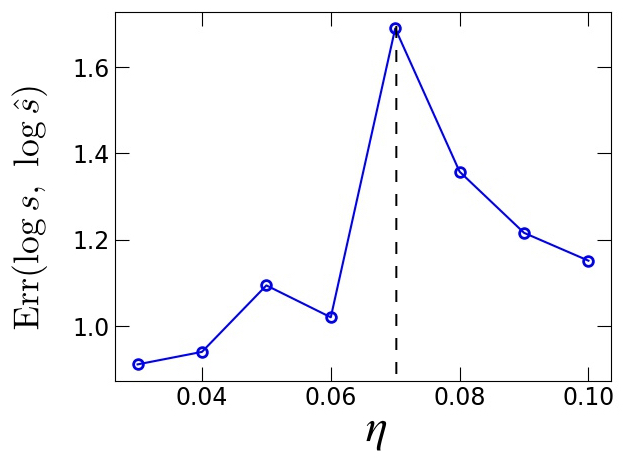}
\caption{The testing error of the predicted event sizes as a function of the noise $\eta$ with $\text{Err}(\log \hat{s},\, \log s) \equiv 1/N\sqrt{\sum_{i=1}^N(\log s_i-\log \hat{s}_i)^2}$. The vertical dashed line indicates the location of the critical noise $\eta_c \approx 0.07$~\cite{Matin_draft}. Note the poorer predictability as the critical point (denoted by the vertical dashed line) is approached.}
\label{fig:MSE_noise}
\end{figure}

In Fig.~\ref{fig:MSE_noise} we plot the testing error $\text{Err}(\log \hat{s},\, \log s) \equiv 1/N\sqrt{\sum_{i=1}^N(\log s_i-\log \hat{s}_i)^2}$ as a function of $\eta$. The reason for using $\log s$ instead of $s$ in the error function is because of the larger fluctuations in $(\hat{s}-s)$ for larger $s$ and because we are interested in the relative error rather than the absolute error. The peak at $\eta_c$ indicates that prediction is not possible in the OFC model at criticality using only the stress grid and the CNN.

\begin{figure}[t]
\centering
\includegraphics[width=7cm]{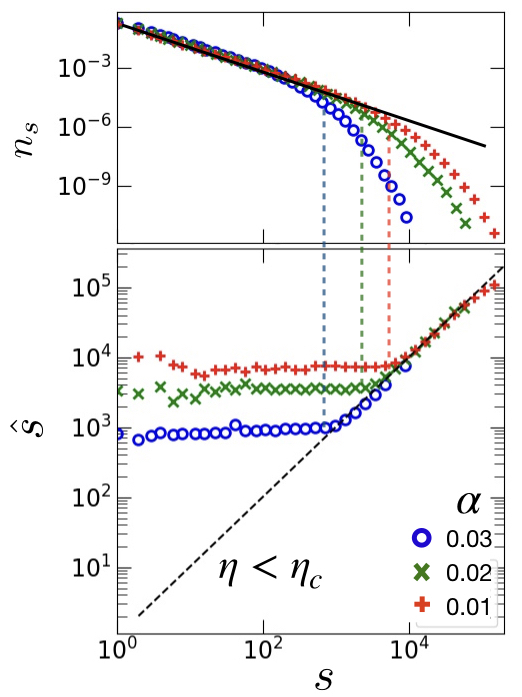}
\caption{Top: the event size distribution $n_s$ for different values of $\alpha$. Bottom: the true values of $s$ versus the predicted $\hat{s}$ event sizes for $\alpha=0.01$ ($\color{red}+$), 0.02 ($\color{green}\times$) and 0.03 ($\color{blue} \circ$). Note that the onset of predictability occurs for $s \gtrsim s_{c,\alpha}$. The vertical dashed line indicates the estimated cutoff $s_{c,\alpha}$ for different values of $\alpha$.}
\label{fig:YpredictYtest_alpha}
\end{figure}

We next look at how the values of the dissipation parameter $\alpha$ affect the predictability of the system as $\alpha \to 0$. No scaling function has been found to fit the dependence of $n_s$ on $\alpha$ in the nearest-neighbor OFC model. Nevertheless, we can determine the cutoff $s_{c,\alpha}$ from $n_s$ as the value of $s$ for which $n_s$ deviates from a power law (see Fig.~\ref{fig:YpredictYtest_alpha}). As $\alpha$ decreases, the cutoff $s_{c,\alpha}$ increases. We observe that the onset of predictability is close to $s_{c,\alpha}$ and the trend persists for different values of $\alpha$.

\section{Visualizing the CNN}
We next explore the  features that the machine has learned which allow it to successfully forecast the size of the non-scaling events and discuss why the critical events are difficult to forecast. We will use occlusion sensitivity analysis to identify the regions of importance of the images that are used by the CNN~\cite{Zeiler_2014}. For example, the face of a dog is expected to contain the most relevant features in determining the type of animal. Hence, blocking the face of the dog should increase the classification error of the CNN. We implement a similar analysis by defining an occluded region in the stress grid and sweeping the occluded region across the entire image to create a map that shows the regions that are the most sensitive to the occlusion. In this way we associate the region that gives the largest change in the predicted event size $\hat{s}$ with the region that is most useful in determining the size of the event.

\begin{figure}[h]
 \centering
\includegraphics[width=14.5cm]{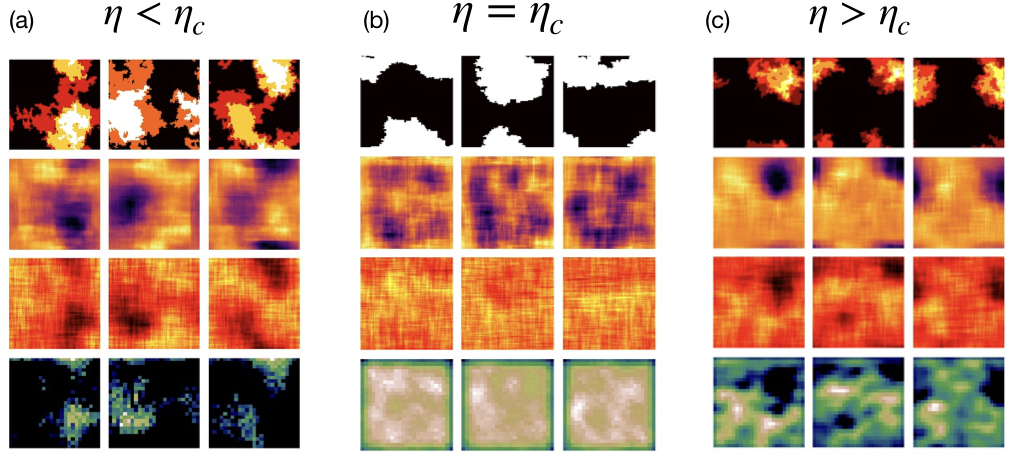}
\caption{Top row: the number of times that a site has failed (failure map). Brighter colors represent more failures. Second row: the sensitivity map from the occlusion sensitivity analysis. Darker regions are more sensitive to the occlusion of that region. Third row: local average stress map. Darker colors represent higher stress. Bottom row: several channels (features) chosen from the third layer in the CNN. Note that the four rows are structurally similar for (a) $\eta<\eta_c$ and (c) $\eta>\eta_c$.}
\label{fig:Visualize}
\end{figure}

In Fig.~\ref{fig:Visualize} we visualize three randomly chosen samples for which $s > s_c$ for (a) $\eta<\eta_c$, (b) $\eta=\eta_c$, and (c) $\eta>\eta_c$. In the top row we show the failure maps, which  correspond to  the number of times that a site has failed. In the second row we show the sensitivity maps from the occlusion sensitivity analysis. Because we chose events of size $s>s_c$, an occlusion that yields a decrease in the predicted event size $\hat{s}$ implies a worse prediction. For $\eta \neq \eta_c$, the region that gives the largest decrease in $\hat{s}$ if occluded coincides with the failure region. We call the region with the largest increase in $\hat{s}$ if occluded the sensitive region. In the third row, we plot the local average stress map. To determine this map, the local average stress of a site is computed by averaging the stress of sites within a square of linear dimension $b=10$, centered at that site. Note that the region of high local stress overlaps with the failure  and  most sensitive regions. This consistency is reasonable because regions with high local stress have a greater probability of initiating and sustaining a large event.

Among the 32 channels in the third layer of the CNN we chose the channel that is visually the most similar to the structure of the failure region.  We interpret the channel as the high level feature learned by the CNN. We plot the channels in the bottom row of Fig.~\ref{fig:Visualize}. From these channels, we see that the machine has learned the connection between the high local stress  and  failure regions.

\begin{figure}[h]
\centering
\includegraphics[width=8.5cm]{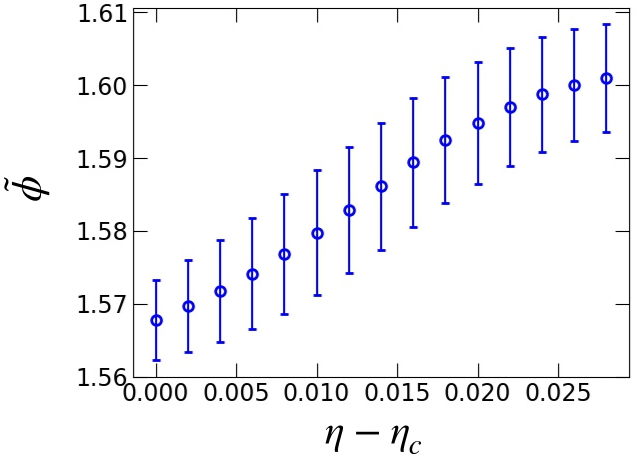}
\caption{The density of the high stress region $\tilde{\phi}$ versus $\eta_c-\eta_c$. We hypothesize that the decrease in predictability at $\eta_c$ is due to the decrease in the density of the high stress region because the CNN has a more difficult time identifying the high stress region.}
\label{fig:Ddensity}
\end{figure}

To understand why prediction is difficult at $\eta=\eta_c$, we look at the failure maps in the top row of Fig.~\ref{fig:Visualize}. We see that the failure regions become more diffuse at $\eta_c$ compared to the more compact failure regions away from $\eta_c$. Although the local average stress map and the failure map remain qualitatively similar, the stress gradient between the high local stress region and the surrounding background is much smaller at $\eta=\eta_c$.

More quantitatively, we define a high stress region as a collection of nearest-neighbor connected sites whose local stress is above the cutoff $\overline{\sigma}_c$ (the bar denotes the stress is the local average stress). We measure the radius of gyration $R_g$ of the largest high stress region in each sample and define the density of the high stress region $\tilde \phi$ as the sum of the local stress within the area of radius $R_g$ divided by $\pi R^2_g$. The density of the high stress region $\tilde \phi$ decreases as $\eta_c$ is approached (see Fig.~\ref{fig:Ddensity}). The smaller density difference  makes it more difficult for the machine to obtain the appropriate cutoff for the high stress region, thus making prediction more difficult. Multiple failures are more prominent for very large events for $\eta\neq\eta_c$, which is why the machine  underestimates the event sizes of the very large events [Figs.~\ref{fig:YpredictYtest}(a) and (c)].

\section{Discussion}
Since we have normalized the stress grid by the average stress before training the machine, the first and second moments of the stress grids do not contain information that can be used by the CNN to predict  event sizes. The fact that the machine learns the association between the region of high local average stress and the event size means that the machine has learned the optimal cutoff that separates the high stress regions from the low stress regions. This task becomes increasingly difficult as the critical point is approached because the high stress region becomes less distinguishable from its background. 

We have found evidence that events whose size distribution satisfies a power law lack distinguishable features which allow the machine to predict their size. This lack of distinguishable features is related to the difficulty of distinguishing between the fluctuations and the background at critical points~\cite{Coniglio_1980}. For the large non-scaling events, there exists  features that allow the machine to successfully predict the event sizes. Similar conclusions are found for both the noise~\cite{Matin_draft} and dissipation~\cite{Christensen_1991} transitions. Our results suggest that large non-scaling events are qualitatively different from the smaller scaling events. This conclusion agrees with the conjecture~\cite{Bak_1989} that prediction is not possible at a true critical point ($L\to\infty$), where there is no deviation from a power law for large events.

It is known that small, large, and very large events in the long-range Rundle-Jackson model~\cite{Rundle_1977} are due to different mechanisms, namely, fluctuations about the spinodal critical point, and failed nucleation and arrested nucleation events, respectively~\cite{Klein_2000, Anghel_2000, Xia}. These different mechanisms suggest that very large events are in principle distinguishable from other events. The caveat is that all three  types of events follow a power law, albeit with different exponents. It will be interesting to see if a machine can learn the difference between the different scaling events.  It is important to note that the failed and  arrested nucleation events, despite the fact that they satisfy a power-law distribution, do not exhibit the same diffusive nature as the smaller events (spinodal fluctuations) on the Gutenburg-Richter scaling plot. This difference appears to be what the CNN picks up.

\section*{Acknowledgement}
We would like to thank Huang Shan and Pankaj Mehta of the Department of Physics at Boston University for useful discussions. 

\appendix*

\section{Sampling method and CNN architecture}
After discarding the transient ($10^6$ plate updates), we run for an additional $10^7$ plate updates and record the event sizes and the random number seed. We then construct the event size distribution and randomly chose five samples from each bin of the distribution (or the number of samples in that bin if there are less than five samples) and record the time of  events in each bin. We then re-run the simulation using the same random number seed and save the stress grids at the recorded times. This procedure ensures that the number of samples in each bin  remains the same for different values of the noise.

The architecture of the CNN consists of 8 alternating layers of convolutional layers and maxpooling layers (4 layers each). The depths of the convolutional layers are 8, 16, 32 and 64, each with a filter of size 5 $\times 5$. We used zero padding on the boundaries to ensure the same size after each convolution. The output of the last maxpooling layer is connected to a fully connected neural network with one hidden layer of 25 nodes. All layers use relu (rectified linear unit) as the activation function except the last layer, which uses a linear activation function. Dropout~\cite{Dropout_2014} with $\mbox{dropout rate} = 0.1$ is applied to the layer immediate before the fully connected layer.


\end{document}